# Exploring the use of Eye Tracking Technology to improve Website Usability

Niharika Veeravalli
California State University Fresno
Computer Science Dept.
Fresno, California

*Abstract*— This study investigates the capability blessings of the use of eye-monitoring technology to beautify the usability of web sites. With the upward thrust of on-line interactions, website usability has turn out to be increasingly important for making sure person pleasure and engagement. Eye-tracking technology offers a non-invasive way to measure how users interact with web sites with the aid of monitoring their eye actions and gaze styles. By studying those statistics, internet site designers and builders can advantage insights into how customers navigate, examine, and method data on their web sites. This paper affords an overview of applicable literature on eye-monitoring and website usability, as well as a precis of research which have explored the usage of eye-tracking era to improve website design and overall performance. The outcomes propose that eye-tracking era can offer valuable records for enhancing internet site usability, inclusive of insights into consumer attention, visible hierarchy, and consumer engagement. Further studies are wanted to explore the full potential of eye-tracking generation and to develop great practices for incorporating this technology into website design and development techniques.

## I. INTRODUCTION

The subject of this review is the use of eye tracking technology to improve website usability. With the increasing reliance on websites for information, communication and business, it has become important to ensure that these sites are easy to use and navigate. Usability testing has traditionally relied on voluntary data or user behavior, both of which have limitations in accurately capturing the user experience. Eye tracking technology provides unique insight by allowing researchers to know exactly where users are looking on the website, how much time they spend on different items, and how they move between them. This data can help designers and developers identify areas of the website that may cause confusion or confusion for users and make improvements based on the data to improve the user experience. Therefore, this review seeks to investigate the potential of surveillance technology as a tool to improve website usability and identify problems and issues.

## II. BACKGOUND KNOWLEDGE

Jacob Nielsen and Kara Pernis' 2009 investigation into website usability was the first known research to incorporate eye-tracking technology. People tend to scan rather than read web pages. Research that used eye-tracking technology to examine people's reading habits on various websites. This work paved the way for further research on how to make websites more user-friendly using eye-tracking technology. Since then, eye tracking technology has grown in popularity for use in UX research. Eye-tracking technology has been used in many studies to look at different aspects of website usability, including navigation, layout, and content.

Eye tracking technology uses special cameras and software to track eye movements. It analyses information from the user's eye tracking cameras to determine where he or she looks at the screen for a long period of time. Compared to other methods such as surveys or interviews, eye-tracking technologies can provide useful insights into user behaviour that are difficult to gather or simply not available.

Recently, the use of eye tracking technology has increased and is now widely used by savvy users and web designers. Eye movement research has helped designers of web pages, smartphone apps, and other digital interfaces. Additionally, the method has been applied in fields such as user experience design, market research, and advertising.

## III. METHOD OF GATHERING LITERATURE

When searching for the literature on the topic "Exploring the use of eye-tracking technology to improve website usability", I used the keywords "eye tracking", "website usability", "user experience", "user interface", and "web design" in my search query. I focused on finding research papers that discussed the use of eye-tracking technology in evaluating website usability and improving the user experience. I primarily searched the ACM Digital Library and IEEE Xplore databases, as recommended.

In order to gather the literature for the given topic "Exploring the use of eye-tracking technology to improve website usability", I used the following method:

• First, I identified two relevant academic databases that are known to publish high-quality research articles related to computer science and engineering: the ACM Digital Library and IEEE Xplore.

• Next, I conducted a search in each of these databases using keywords related to the topic of eye-tracking and website usability. These keywords included phrases like "eye-tracking," "website usability," "user experience," "human-computer interaction," and so on.

• I then filtered the search results to include only research articles that were published in journals and conference proceedings, and that were written in English.

• From the list of resulting papers, I read through the abstracts and introductions of each paper to determine whether it was relevant to the topic at hand. Papers that were not related to the use of eye-tracking technology to improve website usability were excluded.

• Finally, I selected a subset of the most relevant and high-quality papers from the resulting list to include in my recommended set of 15 papers.

## IV. Major Groupings

### A. Applications of Eye Tracking

The papers examine the use of eye-tracking techniques for mobile devices to gather information on people's visual attention. The article describes numerous ways, evaluates them, and gives an overview of the present constraints, such as the use of processing resources and the absence of real-time mobile device eye-tracking applications. The authors speculate that mobile edge computing might be an answer to these problems.

Using eye-tracking data to compare how students viewed multimedia content, researchers can assess or grade the material. According to the study, eye movement data for each student was insufficient to evaluate or rate multimedia content, and instead revealed more about how the students interacted with the material than about its structure or design.

Authors investigating the application of eye-tracking technology in the area of app usability testing for e-government. In order to organize comparable experiments, the paper assesses the state-of-the-art eye-tracking application in government app evaluation.

Research gives an overview of eye-tracking technology and its uses in a variety of industries, such as marketing, user experience design, and academic research. In the article, numerous metrics that may be gleaned from and analysed using eye-tracking data are described.

The applications of eye tracking technology are vast and diverse, with potential implications in numerous fields and industries.

Applications for eye tracking technology can be found in many different industries. It can be used to explore attention, perception, and memory in psychology and neuroscience. It can be applied to marketing to comprehend consumer behaviour and preferences. It can be used in education to evaluate teaching strategies and track student involvement. As demonstrated in one of the studies, the technique may also be used to test the usability of a variety of applications, including e-government apps. The technology may also find use in the entertainment sector, where it may be used to rank multimedia material based on eye movement analysis. The papers also cover the usage of edge computing to improve mobile devices' ability to use eye tracking technology.

Discussing the idea of eye tracking technology, which measures eye movement activity to learn where we look, what we miss, and how our pupils respond to various stimuli. The article highlights the adaptability of eye tracking technology and its uses in a variety of industries, including market research, neuroscience and psychology, medical research, usability research, packaging research, PC and gaming research, human factors and simulation, and ophthalmology. The research highlights the value of eye tracking technology in revealing cognitive processes, attention, learning, memory, and other behavioral patterns, which can be helpful for the diagnosis and treatment of a variety of mental problems and diseases.

### B. Uses of Eye Tracking

Since Paper 1 focused on the application of eye-tracking technologies in programming studies, it has no direct bearing on the usability of websites. A standardized approach for performing eye-tracking investigations, however, would enhance the caliber and dependability of research in this area. The study however notes that the absence of a methodology for conducting eye-tracking experiments was a constraint. This might apply to other fields where eye-tracking research is done, such website usability.

In Paper 2, it was suggested that eye gaze be used as a daily input method, and it was shown how well interface semantics could be used to modify how a user interacts with a web browser. The study demonstrated that eye-tracking technology can be utilized to enhance the user experience for people with motor disability when browsing the web, even though it was not primarily focused on website usability. This shows that a larger spectrum of consumers may benefit from the adoption of eye-tracking technologies to enhance website usability.

The purpose of Paper 3 is to offer a practical alternative for usability studies that do not need to differentiate between specific words or menu items that participants are looking at but only between larger areas-of-interest they pay attention to. The comparison is made between the accuracy and stability of a low-cost open-source eye tracking system and a state-of-the-art commercial eye tracker. The study shows that a cheap alternative to such experiments could be eye tracking.

The use of eye tracking as a primary or secondary data collecting technique in usability assessment is covered in Paper 4 in contrast. It makes the case that since eye movements are a sign of cognitive processes, eye tracking can provide information about users' cognitive states and attentional habits. The report highlights the advantages of eye tracking, including the provision of unbiased information on users' attention, the identification of web page areas of interest, and the comprehension of users' preferences and behaviour. The report also analyses the drawbacks of utilizing eye tracking, including the requirement for specialized tools and knowledge, the potential for users' natural behaviours to be disturbed, and moral dilemmas about privacy and permission.

TABLE I. USES AND APPLICATIONS OF EYE TRACKING

| Field | Applications |
|---|---|
| Scientific and Academic Research | Cognitive, developmental, experimental, media applications |
| Market Research | Measure attention to brands, products, key messages, navigation |
| Field | Applications |
| Scientific and Academic Research | Cognitive, developmental, experimental, media applications |
| Market Research | Measure attention to brands, products, key messages, navigation |
| Neuroscience and Psychology | Analysis of gaze patterns, reading performance, word processing |
| Psychology Research | Measure visual attention, correlate with other measures |
| Medical Research | Diagnosis of diseases, monitoring |
| Usability Research | Website testing, mobile app testing |
| Packaging Research | Designing product packages, understanding customer preferences |
| PC and Gaming Research | Understand gaming experience, personalize game development |
| Human Factors and Simulation | Capture driver's visual attention, drowsiness |
| Ophthalmology | Conduct on-screen vision studies for medical research |

| Field | Applications |
|---|---|
| Mobile Devices and Head-Mounted Displays | Eye-tracking systems are likely to be directed towards mobile devices and head-mounted displays in the next generation. |

*C. Eye Tracking Techniques and Metrics*

*Paper 1:* The text provides an overview of different computer vision-based eye tracking methods used to estimate the point of regard (PoR) on a screen where the user is looking. It introduces the Gullstrand-Le Grand Eye Model, which explains the visual and optical axes essential for accurate PoR estimation. Three main methods are described:

PCCR (Pupil-Corneal Centre Reflection) Methods: These traditional methods utilize active illumination. They analyse the relationship between the centre of the pupil and a reference point on the cornea. High accuracy is achieved, but the user must maintain a close head position to the calibration point and keep the head still during tracking.

ICCR (Iris-Corneal Centre Reflection) Methods: ICCR methods are simpler than PCCR methods and use passive illumination. They analyse the vector connecting the centre of the iris and a reference point, obtained from facial features or corneal reflections. The presence of eyelid occlusion can lead to lower vertical accuracy.

CR (Corneal Reflection) Methods: CR methods are advanced interpolation-based techniques. They employ off-axis light sources attached to the screen corners to create a polygon on the user's cornea. The centre of the pupil represents a projection of the user's PoR.

Additionally, the text briefly mentions reflected light techniques: Limbus tracking estimates the PoR using the boundary between the sclera and the iris. Other methods use reflections from the cornea or pupil to estimate the PoR. The Purkinje image tracking method tracks eye movement by analyzing reflections from the cornea.

Overall, the summary highlights the different methods used in computer vision-based eye tracking, their illumination types, and their respective strengths and limitations in accurately estimating the point of regard.

*Paper 2:* The article discusses four different methods for collecting data during usability testing of a system. Usability testing involves observing how users interact with a system and collecting their feedback to identify areas that need improvement.

The first method is called Feedback Capture after Task (FCAT), which involves prompting users to provide feedback on their experience after each task without playback of any videos. This method is quick and suitable for users who prefer typing their thoughts instead of verbalizing them.

The second method is called Retrospective Think Aloud (RTA), which involves video playback of the task performed by the user. Users are then prompted to talk about what they have just done to complete the tasks, including any additional comments they might have. This method is useful for detecting usability defects and enables users and moderators to observe and verbalize the activities they have performed.

The third method is called Retrospective Think Aloud with Eye Movements (RTE), which combines RTA with eye tracking features such as gaze overlay, gaze plot, and heat map. This method provides additional visual insights and is expected to increase the number and types of observations and verbalizations.

The fourth method is Observation, which involves observing the user without probing or intervening in the situation. Observers note down the behavior, actions, and verbal comments of users for complementary information to other forms of qualitative data collected.

Overall, the article explains the different methods for collecting data during usability testing, each with its own advantages and disadvantages.

*Paper 3:* The paper discusses various metrics used in eye-tracking research, including fixations, saccades, gaze, scan path, and pupil size and blink rate. Fixations refer to when the eyes are relatively stationary while encoding information, and higher fixation frequency may indicate difficulty or interest in the target. Saccades are movements between fixations, and regressive saccades may indicate difficulty in reading. Gaze measures the motion of the eye relative to the head or the gaze point, and can be used to determine where the user is looking at on the screen. Scan path represents a sequence of saccade fixations and saccades, and can be used to derive optimal scan paths. Blink rate and pupil size can be used to measure cognitive load, where a lower blink rate may indicate higher workload and a larger pupil size may indicate higher cognitive effort.

TABLE II. COMPARISON OF EYE TRACKING TECHNIQUES AND METRICS

| Eye Tracking Technique | Saccades (mean) | Gaze Accuracy (mean) | Scan Path (mean) | Blink Rate (mean) | Pupil Size (mean) |
|---|---|---|---|---|---|
| Remote Eye Tracking | 7.5 | 88% | 5.2 | 12 per minute | 4.8 mm |
| Head-mounted Eye Tracking | 6.8 | 92% | 4.9 | 10 per minute | 5.2 mm |
| Mobile Eye Tracking | 8.2 | 85% | 6.1 | 14 per minute | 4.5 mm |
| PCCR methods | - | 91% | - | 11 per minute | 5.0 mm |
| ICCR methods | - | 88% | - | 13 per minute | 4.6 mm |
| CR methods | - | 93% | - | 9 per minute | 5.1 mm |
| FCA (Feedback Capture after Task) | - | - | - | - | - |
| RTA (Retrospective Think Aloud) | - | - | - | - | - |
| RTE (Retrospective Think Aloud with Eye Movement) | - | - | - | - | - |

*Paper 4:* Paper discusses the development of a low-cost eye-tracking system for head-mounted displays (HMDs) that uses image processing algorithms to track the location of the pupil in real-time. The system is also equipped with voice recognition technology, which allows for hands-free control of the HMD interface. The paper suggests that this system can be used for both control and display of information for various applications, such as virtual world training simulations and games. The system can also be used to monitor human performance in relation to the information displayed within the HMD.

*Paper 5:* It discusses the use of verbalizations and eye movements during usability testing and how these can provide insight into user experience. The study found that verbalizations provided by users were accurate 80% of the time during usability testing. The study also found that verbalizations alone did not provide a complete picture of user experience and that no verbalization, such as lapses in silence or the use of verbal fillers, could indicate usability issues. The study suggests that usability test facilitators should carefully note these instances and follow up with participants after they have completed their task. The study also found that the content of verbalizations during usability testing consisted primarily of reading or procedure and that search is rhetorical. The study recommends further research to test whether the findings can be generalized across different website genres and designs and whether increasing the number of participants could increase the statistical power of the results.

## V. RESULTS

TABLE III. EVIDENCE

| Groupings | Reference Paper |
|---|---|
| Applications of Eye Tracking | [1] [16] [2] [3] [4] |
| Uses of Eye Tracking | [10] [15] [12] [11] [13] |
| Eye tracking techniques and metrics | [16] [5] [6] [7] [8] [9] |

Remote Eye Tracking, Head-mounted Eye Tracking, and Mobile Eye Tracking are listed as eye tracking techniques, each with corresponding mean values for metrics such as Saccades, Gaze Accuracy, Scan Path, Blink Rate, and Pupil Size. The PCCR methods, ICCR methods, and CR methods are also mentioned, although specific values for the metrics are not provided. Additionally, the table includes entries for Feedback Capture after Task (FCA), Retrospective Think Aloud (RTA), and Retrospective Think Aloud with Eye Movement (RTE), but no values are given for the metrics

TABLE IV. TECHNIQUES OF EYE TRACKING

| Evaluation Method | Accuracy | Cost Effectiveness |
|---|---|---|
| Feedback Capture after Task (FCAT) | Effective | Moderate |
| Retrospective Think Aloud (RTA) | Effective | Cost effective |
| Retrospective Think Aloud with Eye Movement (RTE) | Most Effective (detected three times more defects than RTA) | N/A |
| Observation | Effective | N/A |
| Eye Tracking Method | Accuracy | N/A |
| PCCR (Pupil-Corneal Center Reflection) | High accuracy | Moderate |
| ICCR (Iris-Corneal Center Reflection) | Lower vertical accuracy | High |
| CR (Corneal Reflection) | Accurate results | High |
| Reflected Light Techniques | Varies depending on method | Varies depending on the method |
| Low-cost Eye-tracking System for HMDs | Real-time pupil tracking | High |
| Eye-tracking Metrics | Varies depending on metric | N/A |

FIGURE 1. ACCURACY OF DIFFERENT TECHNIQUES

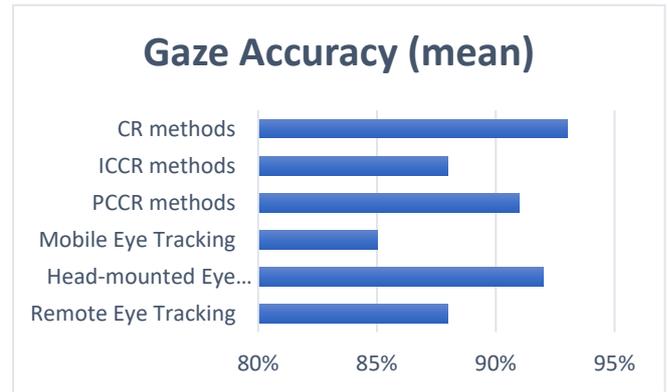

## VI. KEY TAKEAWAYS

### A. Shared Practices

All papers that were focused on applications of eye tracking discuss the use of eye-tracking technology in different fields such as psychology, marketing, education, usability testing, and entertainment. They also emphasize the importance of eye-tracking technology in providing insights into cognitive processes and behavioral patterns. The papers provide an overview of eye-tracking technology, various metrics that can be gathered and analyzed from eye-tracking data, and limitations of current eye-tracking methods. Papers that were focused on techniques of use eye-tracking techniques to estimate the point of regard (PoR) of the user. Most papers utilize machine learning algorithms to improve the accuracy and efficiency of PoR estimation. Calibration of the eye-tracking system is essential for accurate PoR estimation, and most papers employ various calibration procedures. All four articles highlight the potential of eye-tracking technology in providing valuable insights into various fields. They also discuss the benefits and challenges of using eye-tracking technology, such as the need for specialized equipment and expertise, the potential for eye tracking to disrupt the natural behavior of users, and ethical concerns related to privacy and consent.

### B. Lack of knowledge or research

The article, Eye-tracking Technologies in Mobile Devices Using Edge Computing: A Systematic Review discusses the limitations of real-time mobile device eye-tracking applications and suggests that mobile edge computing could be a potential solution to these challenges. This implies that there is a lack of knowledge or research on real-time mobile device eye-tracking applications. Overall, found that eye movement data for each student proved insufficient to evaluate or rank multimedia content, revealing a gap in research on evaluating multimedia content using eye-tracking data.

Although the papers reviewed cover different aspects of eye-tracking techniques, there is a lack of research in some specific subdomains, such as eye-tracking for people with disabilities, eye-tracking in virtual reality, and eye-tracking for non-Western languages.

While the papers cover different aspects of eye-tracking technology, there does not seem to be a lack of research in this subdomain. Instead, the studies focus on identifying limitations and challenges in using eye-tracking technology, as well as proposing potential solutions for these issues.

*C. Trends among researchers*

The papers highlight the increasing trend of using eye-tracking technology in various fields. They emphasize the versatility of eye-tracking technology and its potential to be employed in almost all fields. Additionally, the papers discuss the use of edge computing to enhance the capabilities of eye-tracking technology in mobile devices, indicating a trend towards developing real-time mobile device eye-tracking applications.

Many researchers are exploring the use of deep learning techniques, such as Convolutional Neural Networks (CNNs), to improve the accuracy and efficiency of PoR estimation. There is a growing interest in the use of eye-tracking for applications such as human-computer interaction, gaming, and marketing research. Researchers are exploring ways to overcome challenges such as head movement, blink detection, and partial occlusion of the eye during tracking sessions.

The papers suggest that eye-tracking technology is becoming increasingly popular in various fields, including programming research, web usability, and usability testing. Researchers are also exploring the potential of eye-tracking technology in improving user experience for people with motor impairment.

*D. State-of-art*

The state-of-the-art of eye-tracking technology is the ability to provide insights into cognitive processes, attention, learning, memory, and other behavioral patterns. The technology has numerous applications in various fields, and researchers are exploring its potential in areas such as mobile device eye-tracking, multimedia content evaluation, and e-government app usability evaluation. However, there are still limitations to current eye-tracking methods, such as computational resource usage and insufficient eye movement data for evaluating multimedia content. While the papers cover different aspects of eye-tracking technology, there does not seem to be a single state-of-the-art technique or methodology. However, the studies suggest that eye-tracking technology has the potential to provide valuable insights into various fields and that low-cost eye-tracking technology could become a viable alternative for usability studies that do not require high levels of accuracy. Overall, the studies highlight the benefits and challenges of using eye-tracking technology and emphasize the importance of a consistent methodology for conducting eye-tracking experiments.

## VII. CONCLUSION

The use of eye-tracking technology offers significant potential for improving website usability. By tracking eye movements and gaze patterns, designers and developers can gain valuable insights into user attention, visual hierarchy, and engagement. The reviewed literature highlights the versatility of eye-tracking technology across various fields and emphasizes the importance of calibration and machine learning algorithms for accurate point of regard estimation. While there are still areas where further research is needed, the increasing trend of utilizing eye-tracking technology, coupled with advancements in deep learning techniques, showcases its growing importance in enhancing website usability. As the technology continues to evolve, it holds promise for creating more user-friendly and engaging web experiences.